\documentclass[preprint,aps,showpacs,showkeys,amsmath,amssymb]{revtex4}

\begin{document}

\title{On the number-phase problem}
\author{S. Dumitru}
\date{\today}
\email{s.dumitru@unitbv.ro}
\affiliation{Department of Physics, ``Transilvania'' University, 
B-dul Eroilor 29, R-2200, Bra\c sov, Romania} 

\begin{abstract}
The known approaches of number-phase problem (for a quantum oscillator) 
are mutually contradictory. All of them are subsequent in respect with the 
Robertson-Schr\"{o}dinger uncertainty relation (RSUR). In opposition here it is 
proposed a new aproacch aimed to be ab origine as regard RSRUR. 
From the new perspective the Dirac's operators for vibrational number 
and phase appear as correct mathematical tools while the alluded problem 
receives a natural solution. 
\end{abstract}


\maketitle

\section{Introduction}
A recent outstanding work \cite{1} reviews the publications
referring to the problem of theoretical description for the
vibrational number $N$ and phase $\Phi$ of a quantum oscillator (QO). 
So one discloses the fact that the respective problem known various
approaches differing among them both quantitatively and
qualitatively. Moreover it is pointed out that until now in scientific
literature an agreement regarding the mentioned problem does not
exist. As we shall show below all the alluded approaches are
subsequent in respect with the Robertson-Schr\"{o}dinger uncertainty
relation (RSUR). In such a context we think that another approach,
ab origine regarding RSRUR, as the one which we present in the next
sections, can be of nontrivial interest.

\section{Briefly on known facts} 
The story of $N-\Phi$ problem \cite{1} started with the
Dirac's idea to transcribe the ladder (annihilation and creation)
operators $\hat{a}$ and $\hat{a}^+$ in the forms
\begin{equation}\label{eq:1}
\hat{a} = e^{i\hat\Phi} \sqrt{\hat N} \qquad
\hat{a}^+ = \sqrt{\hat N} e^{-i\hat\Phi} 
\end{equation}
For an oscilator $\hat N$ and $\hat\Phi$ were identified with the operators of
vibrational number respectively of phase. Due to the fact that
$[ \hat a, \hat a^+]= \hat a \hat a^+ -\hat a^+ \hat a = 1$ 
from \eqref{eq:1} it follows
\begin{equation}\label{eq:2}
[\hat N , \hat\Phi] = i
\end{equation}

Then this relation was regarded in connection with RSUR
\begin{equation}\label{eq:3}
\Delta A \cdot \Delta B \geq \frac{1}{2} \vert \langle [ \hat A,\hat B ] 
\rangle \vert
\end{equation}
In \eqref{eq:3} the standard deviations $\Delta A$ and $\Delta B$ 
respectively the operators $A$ and $B$ refer to the two arbitrary observables 
$A$ and $B$. By a direct application of \eqref{eq:3} to the $N-\Phi$ case 
it was introduced the relation
\begin{equation}\label{eq:4}
\Delta N \cdot \Delta\Phi \geq \frac{1}{2}
\end{equation}
But, lately it was found that relation \eqref{eq:4} is false - at least in
some well-specified situations. Such a situation appears in the case
of QO eigenstate corresponsding to the energy eigenvalue 
$E_n = \hbar\omega ( n+ \frac{1}{2})$. 
The respective state is described by the wave function $\Psi_n$ for
which one obtains $\hat N\Psi_n = \hat a^+\hat a \Psi_n = n\Psi_n$, 
$\Delta N =0$ respectively $\Delta\Phi \leq 2\pi$
(the noted restriction for the value of $\Delta\Phi$ results \cite{1} \
from the fact that the range of definition for $\Phi$ is the interval 
$[0, 2\pi)$). With the mentioned falsity of \eqref{eq:4} the $N-\Phi$ 
problem reached a deadlock. For avoiding the mentioned deadlock in literature 
various alternative approaches were promoted (see \cite{1} and references). 
But it is easy to emark that all the alluded approaches are subsequent 
(and dependent) in respect with the RSUR \eqref{eq:3} in the following sense. 
The respective approaches consider \eqref{eq:3} as an absolutely valid 
formula and try to adjust accordingly the description of the pair 
$N-\Phi$ for QO. So the operators $\hat N$ and $\hat\Phi$ defined in \eqref{eq:1} 
were replaced by some alternative operators $\hat N_a$ and $\hat\Phi_a$ 
whose standard deviations $\Delta N_a$ and $\Delta\Phi_a$ satisfy
relations resembling more or less with \eqref{eq:3}. 
But it is very doubtful that the variables $N_a$ and $\Phi_a$ have a natural 
(or even useful) physical significance. Probably that this is one of the reasons 
why until now, in scientific community, it does not exist an agreement 
regarding the $N-\Phi$ problem.

\section{An ab origine approach} 
In contrast with the known approaches of $N-\Phi$ problem alluded above 
we think that the same problem can be approached on a new way which is 
ab origine (i.e. non-subsequent) in respect with the RSUR \eqref{eq:3}. 
Such a new approach can be done by investigating the true origin of the relation 
\eqref{eq:3} as well as the conditions of validity for the respective relation. 
For puting in practice our thinking we act as follows. Firstly let us remind some
elements/notations from quantum mechanics. We consider a system (particularly 
an oscilator) of quantum nature. The state of the system and its observables 
$A_j (j=1,2,\ldots,r)$ are described by the wave function $\Psi$ respectively 
by the operators $\hat A_j$. The scalar product of two functions $\Psi_\alpha$  
and $\Psi_\beta$ will be denoted by $(\Psi_\alpha,\Psi_\beta)$. In the state
described by $\Psi$ the expected (mean) value of the quantity $A_j$ is given
by $\langle A_j \rangle= (\Psi,\hat A_j\Psi)$ and the operator
$\delta\hat A_j = \hat A_j -\langle A_j\rangle$ can be defined. 
Then for two osevables $A_1=A$ and $A_2=B$ one can write the following Schwartz 
relation
\begin{equation}\label{eq:5}
(\delta \hat A\Psi, \delta \hat A\Psi) \, (\delta\hat B\Psi, \delta\hat B\Psi)
\geq \vert (\delta\hat A \Psi,\delta \hat B \Psi) \vert^2
\end{equation}

But $(\delta\hat A\Psi,\delta\hat A\Psi)=(\Delta A)^2$ where $\Delta A$ 
denotes the standard deviation of $A$. So from \eqref{eq:5} one obtains
\begin{equation}\label{eq:6}
\Delta A \cdot \Delta B \geq \vert ( \delta\hat A \Psi, \delta \hat B\Psi)\vert
\end{equation}
Note that the relation \eqref{eq:6} is generally valid for any wave function
$\Psi$ and any observables $A$ and $B$. The respective relation imply the
less general formula which is RSUR \eqref{eq:3} only in particular
circumstances. The alluded circumstances can be specified as follows. 
If in respect with the wave function $\Psi$ the operators $\hat A=\hat A_1$
and $\hat B=\hat A_2$ satisfy the conditions
\begin{equation}\label{eq:7}
(\hat A_j \Psi, \hat A_k \Psi)=(\Psi, \hat A_j \hat A_k \Psi) \qquad
(j=1,2;\, k=1,2)
\end{equation}
one can write
\begin{equation}\label{eq:8}
(\delta \hat A \Psi, \delta \hat B \Psi)=\frac{1}{2}
(\Psi, \{ \delta \hat A, \delta\hat B\}\Psi) - \frac{i}{2}
(\Psi, i[\hat A , \hat B]\Psi)
\end{equation}
where both $(\Psi, \{ \delta\hat A,\delta\hat B\}\Psi) = (\Psi,(\delta\hat A
\, \delta\hat B +\delta\hat B\, \delta\hat A)\Psi)$ and $(\Psi,i[\hat A, \hat B]\Psi)$
are real quantities. This means that in the circumstances strictly
delimited by the conditions \eqref{eq:7} the relation \eqref{eq:6} 
imply directly the RSUR \eqref{eq:3}. In all other circumstances 
RSUR \eqref{eq:3} is false but the relation \eqref{eq:6} remains 
always valid.  The above-presented considerations, regarded in connection 
with the here investigated $N-\Phi$ problem, justify the following observations 
(\textbf{Ob}):
\begin{itemize}
\item[\textbf{Ob.1:}] For a state described by an eigenfunction $\Psi_n$ 
mentioned in Sec.2, by using the formula $\hat N\Psi_n = n\Psi_n$ together 
with the relation \eqref{eq:2}, it results
\begin{equation}\label{eq:9}
(\hat N\Psi_n,\hat\Phi\Psi_n) = (\Psi_n, \hat N\hat\Phi\Psi_n)+i
\end{equation}
Such a result clearly shows that in the considered situation $N$ and $\Phi$
do not satisfy the conditions \eqref{eq:7}. This means that, in the
respective situation, for $N$ and $\Phi$ the RSUR \eqref{eq:6} and 
(consequently) the formula \eqref{eq:4} are not valid.  However in the 
same situation the relation \eqref{eq:6} remain true. But then 
$\delta\hat N \Psi_n=0$, $\Delta N = 0$ and $\Delta\Phi \leq 2\pi$
(more exactly $\Delta\Phi = \pi/\sqrt{3}$ - see below \eqref{eq:A.4} in Appendix).
So in the alluded state \eqref{eq:6} degenerates into trivial equality 
$0 = 0$.
\item[\textbf{Ob.2:}]: The cases when the state of the oscilator is described 
by a wave function of the form $\Psi=\sum_n C_n\Psi_n$ ( with 
$\sum_n \vert C_n\vert^2 =1$) must be discussed distinctly. 
Depending on the values of coefficients $C_n$ in respect with such a function 
$\hat N$ and $\hat\Phi$ defined by \eqref{eq:1} can satisfy the
conditions \eqref{eq:7}. Then relation \eqref{eq:4} is valid with 
\begin{equation}\label{eq:10}
\Delta N= \Big[\sum_n \vert C_n\vert^2 n^2 - (\sum_n \vert 
C_n\vert^2 n)^2\Big]^{1/2} \, , \quad \Delta\Phi \leq 2\pi
\end{equation} 
\item[\textbf{Ob.3:}] By means of the operators $\hat N$ and $\hat\Phi$ 
from \eqref{eq:1} can be composed a simple procedure able to reveal 
the important characteristics of a QO (see below the Appendix).
\end{itemize}

\section*{Conclusions} 
In scientific literarure the $N-\Phi$ problem persists as a
controversial question. In the main the controversies originate from
the fact that the operators $\hat N$ and $\hat\Phi$ defined by \eqref{eq:1} 
are incompatible with the RSUR \eqref{eq:3}/\eqref{eq:4}. 
The known approaches of the problem are subsequent in the respect with RSUR 
\eqref{eq:3}. In oposition we propose a new approach which is ab origine 
in relation with RSUR \eqref{eq:3}. Within the framework of the proposed 
approach we point out the delimitative conditions in which the RSUR 
\eqref{eq:3} is valid. Also we noted that in fact RSUR \eqref{eq:3} 
originates from a more general formula namely from the Schwartz relation 
\eqref{eq:6}/\eqref{eq:5} which is ab origine and always valid. 
Then as it is shown in \eqref{eq:9} in the cases of energy eigenstates
of a QO the operators $\hat N$ and $\hat\Phi$ do not satisfy 
the conditions of type \eqref{eq:7} required by RSUR \eqref{eq:3}. 
However in the respective  cases the pair $N-\Phi$ 
satisfies the Schwartz's relation of type \eqref{eq:6} 
which reduces to the trivial equality $0 = 0$. As we have shown in \textbf{Ob.1} 
the mentioned facts give a natural elucidation of the known incompatibility (and
resulting troubles) existing between the Dirac's operators $\hat N$ and $\hat\Phi$
from \eqref{eq:1} and the RSUR \eqref{eq:3}/\eqref{eq:4}. 
In direct connection with the respective elucidation in \textbf{Ob.2} 
we indicate a situation when the Dirac's operators $N$ and $\Phi$ satisfy RSUR 
\eqref{eq:3}/\eqref{eq:4}. Additionally in \textbf{Ob.3} we noted 
that the Dirac's operators $\hat N$ and $\hat\Phi$ prove themselves 
to be useful tools  for composing a simple mathematical procedure 
able to reveal the important characteristics of a QO 
The resulting conclusion of the above discussions is that that the 
ab origine approach presented here gives a complete and natural 
solution to the controversial $N-\Phi$ problem. Accordingly the Dirac's operators 
$\hat N$ and $\hat\Phi$ [defined in \eqref{eq:1}] remains as correct 
thoretical concepts with clear significance and utility for physics. 
From the perspective of the mentioned approach, in the case of a QO, 
the appeals to other ``alternative operators for number and phase'' 
seem to be pure mathematical exercices without major significance for physics.

\appendix*

\section{The oscillator equation in $\Phi$-represen\-tation}
In obtaining the characteristics of a QO the usual procedures
operate with the wave function $\Psi$ taken in x-representation - i.e. 
$\Psi=\Psi(x)$ ($x$ = coordinate). Mathematically the respective procedures 
imply relative lablaborious work (for a direct solution of QO Schr\"{o}dinger
equation or, equivalently, for the iterative handling of the ladder
operators $\hat a$ and $\hat a^+$ in the same equation). 
A more simple procedure can be done if the wave function $\Psi$ taken in 
$\Phi$-representation - i.e. $\Psi=\Psi(\Phi)$ [with $\Phi$ from \eqref{eq:1}]. 
In such a representation from \eqref{eq:2} it results directly that $\hat N$ 
has  the form $\hat N=(i\partial/\partial\Phi)$. Then for the QO
Hamiltonian $\hat H =(\hat p^2/2m)+(m\omega^2\hat x^2/2)=(\hbar\omega/2)
(\hat a\hat a^+ + \hat a^+\hat a)$ one obtains
\begin{equation}\label{eq:A.1}
\hat H = \hbar\omega \left(\hat N +\frac{1}{2}\right)
= \hbar\omega \left( i\frac{\partial}{\partial\Phi}+ \frac{1}{2}\right)
\end{equation}
Consequently the Schr\"{o}dinger equation for QO becomes
\begin{equation}\label{eq:A.2}
\hbar\omega \left( i\frac{\partial}{\partial\Phi} + \frac{1}{2}\right)
\Psi = E\Psi
\end{equation}
From this equation it is very easy to infer the results
\begin{equation}\label{eq:A.3}
\Psi(\Phi)=\Psi_n(\Phi) = C e^{-in\Phi} \, , \qquad
E=E_n=\hbar\omega \left(n+\frac{1}{2}\right)
\end{equation}
The integration constants $C$ and $n$ can be precised as follows. 
The condition $\Psi_n(0) = \lim_{\Phi\to 2\pi} \Psi_n(\Phi)$ requires for 
$n$ to be an integer number. From normalisation condition $(\Psi_n,\Psi_n) = 1$ 
it result $C=1/\sqrt{2\pi}$. Then requiring that, similarly with the case of 
classical oscillator, to have $E > 0$ one finds that $n \geq 0$. 
With the wave function $\Psi_n(\Phi)$ determined as above can be evaluated 
the characteristics of the QO. So for the states described by $\psi_n (\Phi)$
given by \eqref{eq:A.3} (which regard just the situations debated in literature \cite{1})
one obtains
\begin{equation}\label{eq:A.4}
\Delta N =0 \, , \quad \Delta\Phi = \pi/\sqrt{3} \, , \quad
(\delta\hat{N}\psi, \delta\hat{\Phi}\psi)=0
\end{equation}

\bibliography{articol-prl}

\end{document}